\documentclass[%
 reprint,
 superscriptaddress,
 amsmath,amssymb,
 aps, pra
]{revtex4-2}

\usepackage{graphicx}% Include figure files
\usepackage{dcolumn}% Align table columns on decimal point
\usepackage{bm}% bold math
\usepackage{physics} 
\usepackage{xcolor}
\usepackage{textcomp}
\begin{document}

\preprint{APS/123-QED}

\title{The design of an experimental platform for hybridization of atomic and superconducting quantum systems}

\author{Alessandro Landra}
\affiliation{%
 Centre for Quantum Technologies, National University of Singapore, 3 Science Drive 2, Singapore 117543, Singapore
 }%
\author{Christoph Hufnagel}
\affiliation{%
 Centre for Quantum Technologies, National University of Singapore, 3 Science Drive 2, Singapore 117543, Singapore
 }%
\author{Lim Chin Chean}
\affiliation{%
 Centre for Quantum Technologies, National University of Singapore, 3 Science Drive 2, Singapore 117543, Singapore
 }%
 \author{Thomas Weigner}
\affiliation{%
 Centre for Quantum Technologies, National University of Singapore, 3 Science Drive 2, Singapore 117543, Singapore
 }%
 \affiliation{%
 Division of Physics and Applied Physics, Nanyang Technological University, 21 Nanyang Link, Singapore 637371, Singapore
 }%
 \affiliation{%
 Atominstitut TU Wien, Stadionallee 2, Vienna 1020, Austria
 }%11
 \author{Yung Szen Yap}
\affiliation{%
 Centre for Quantum Technologies, National University of Singapore, 3 Science Drive 2, Singapore 117543, Singapore
 }%
 \affiliation{%
Faculty of Science and Centre for Sustainable Nanomaterials (CSNano), Universiti Teknologi Malaysia, 81310 UTM Johor Bahru, Johor, Malaysia
 }%
  \author{Long Hoang Nguyen}
\affiliation{%
 Division of Physics and Applied Physics, Nanyang Technological University, 21 Nanyang Link, Singapore 637371, Singapore
 }%
\author{Rainer Dumke}
\email{rdumke@ntu.edu.sg}
\affiliation{%
 Centre for Quantum Technologies, National University of Singapore, 3 Science Drive 2, Singapore 117543, Singapore
 }%
 \affiliation{%
 Division of Physics and Applied Physics, Nanyang Technological University, 21 Nanyang Link, Singapore 637371, Singapore
 }%

\date{\today}

\begin{abstract}
Hybrid quantum systems have the potential of mitigating current challenges in developing a scalable quantum computer. Of particular interest is the hybridization between atomic and superconducting qubits. We demonstrate a novel experimental setup for transferring and trapping ultracold atoms inside a millikelvin cryogenic environment, where interactions between atomic and superconducting qubits may be established, paving the way for hybrid quantum systems. $^{87}\text{Rb}$ atoms are prepared in a conventional magneto-optical trap and transported via a magnetic conveyor belt into a UHV compatible dilution refrigerator with optical access. We store $5\times10^{8}$ atoms with a lifetime of 794 seconds in the vicinity of the millikelvin stage.
\end{abstract}

\maketitle

%\tableofcontents

\section{\label{sec:introduction}Introduction}

Quantum technologies promise a new era in many existing applications, like computation \cite{Yamamoto2003,Leibfried2003,DiCarlo2009,PhysRevLett.106.130506,Lucero2012,Veldhorst2015}, analog and digital quantum simulation \cite{Kim2010,Simon2011,Blatt2012,Bloch2012,Houck2012,RevModPhys.86.153,Barends2015}, communication \cite{Duan2001,Ursin2007,Gisin2007} and sensing \cite{RevModPhys.89.035002}. Today, quantum states have been realized in many different physical systems like atoms \cite{RevModPhys.82.2313}, solid-states \cite{Devoret1169} or photonic devices \cite{O'Brien2009}. An important step in future quantum technologies is the creation of coherent interfaces between these systems, which will benefit from the combined advantages of each quantum system in an integrated device \cite{Schoelkopf2008,1402-4896-2009-T137-014001,RevModPhys.85.623}. 

A particularly interesting hybrid quantum device is the combination of superconducting (SC) quantum circuits and ultracold atoms. Superconducting qubits \cite{Devoret1169} are widely considered to be among the most mature approaches for quantum computing. They are robust, flexible in design and admit a fast processing of quantum states. Moreover, continuous improvements over the last decade pushed them into the error correction regime \cite{barends_superconducting_2014}. Yet, there are still many challenges that SC qubits are facing. Overall, coherence times of SC qubits are still limited and highly dependent on external factors like fabrication techniques \cite{PhysRevLett.121.090502}. This restricts their use as effective quantum memories. Another obstacle that SC qubits face is that they cannot be interfaced to optical photons directly, which sets severe limits on their long distance networking capabilities \cite{Wendin_2017}.  
Neutral atom qubits are able to compensate for most of these shortcomings. Coherence times of neutral atoms, particularly in cryogenic environments, are on the order of 10 s \cite{NatCommun4.2380}, about three orders of magnitude longer than in SC qubits. In addition, the states of atomic qubits can be transferred directly to optical photons \cite{PhysRevLett.78.3221}. Few experimental realizations for ultracold atoms in cryogenic environment have been reported, however they were limited to 4K and above base temperatures \cite{Hattermann2017,Brandl2016,Minniberger2014,PhysRevLett.98.260407,PhysRevA.80.061604,Emmert2009, PhysRevA.85.041403, Pagano_2018}, which is not suitable for SC qubits integration. 
Here, we describe the realization of an experimental setup for studying hybrid quantum systems, made of ultracold atoms and superconducting circuits. In this hybrid system we can exploit the long coherence of the atomic states and the fast, high fidelity, driving of logical operations of superconducting qubits. Moreover, with a suitable state preparation of the former and a careful engineering of the electromagnetic modes in the latter \cite{PhysRevA.95.053855, 2058-9565-3-4-045007}, we can investigate different coupling mechanisms. Theoretical protocols for this hybrid system have been proposed. The state transfer and CNOT gate operation can be performed between the two species of qubit, while more complex experiments will allow the stabilization of the Rabi oscillation of SC qubit, using atomic clock techniques with feedback control and quantum random access memory \cite{2058-9565-2-3-035005,PhysRevA.95.053811,PhysRevA.94.062301,Yu2016,PhysRevA.93.042329,FischerIOP,FischerA,FischerL}. Two coupling scenarios are available. First is the collective coupling of the Rubidium 87 clock states, $\ket{5^2 S_{1/2},F=1}$ and $\ket{5^2 S_{1/2},F=2}$ to a microwave cavity field at 6.835 GHz, which is simultaneously coupled to a SC qubit \cite{doi:10.1063/1.3651466, PhysRevA.85.020302}. Similarly, the atoms can be excited to a Rydberg state with high principal quantum number, resulting in a strong electrical dipole coupling with the electric fringe field of a planar circuit \cite{PhysRevA.97.053813,1367-2630-20-2-023031}, e.g. the capacitor of a charge or transmon qubit.

\begin{figure}[t!]
  \includegraphics[scale=0.57]{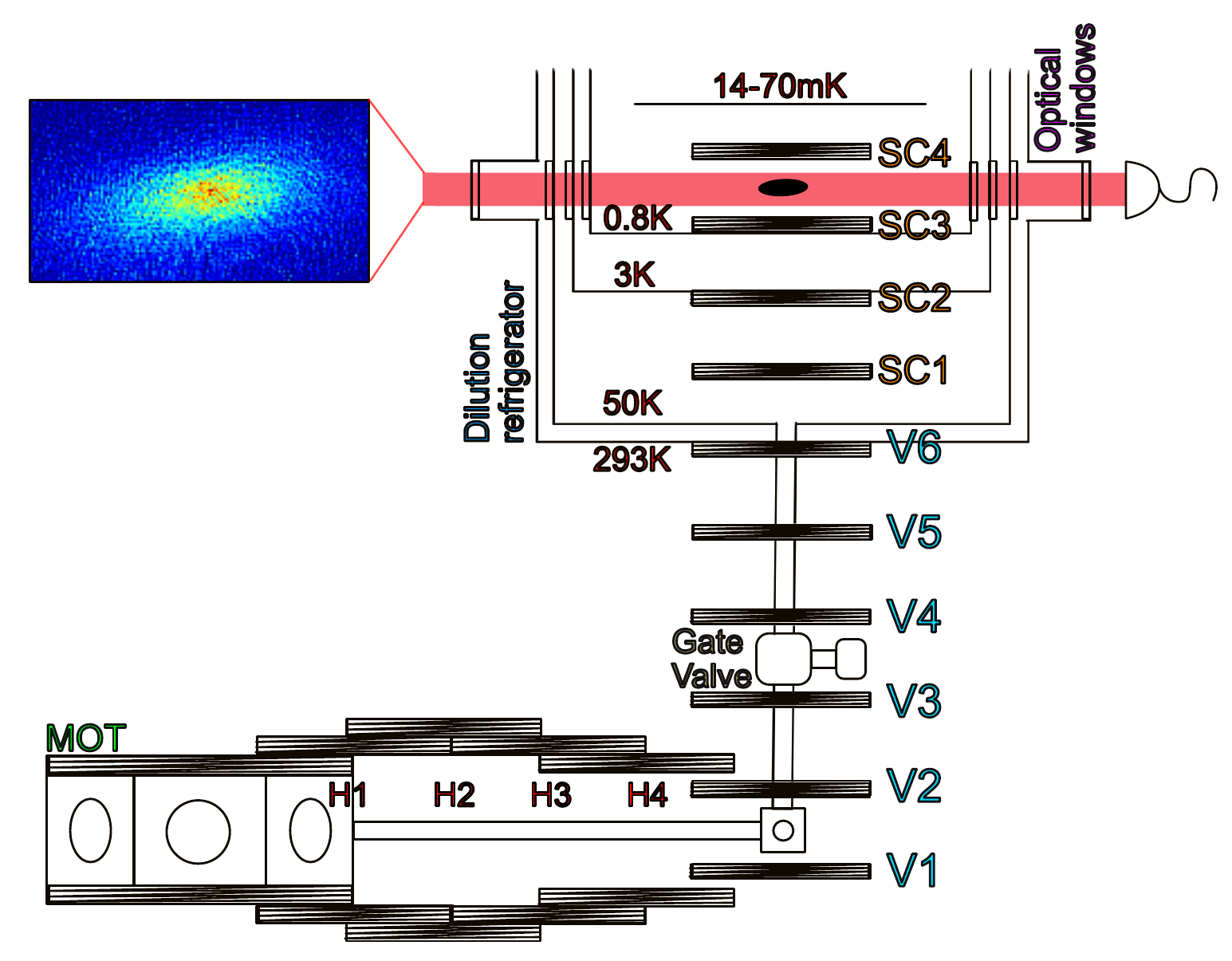}
  \caption{Side view of the experimental setup. The atoms are loaded in the MOT through a Zeeman slower (missing in the schematics), from which a CF16 tube extends on the right side, called horizontal transport (H coils), at the end of which the vertical transport (V coils) starts, up to the cryostat (SC coils). Example of absorption imaging in the inset. The image is acquired through optical windows, available on two perpendicular axes.}
  \label{fig:setup}
\end{figure}

One of the technical challenges in realizing such a hybrid system is the effect of the environment on the  superconducting qubits. Their coherence can be shortened or even destroyed when they are affected by stray magnetic fields or light, meaning that we need to carefully control and shield both factors when and where necessary.

In this report we describe the experimental realization of a platform, suitable for the hybridization of SC and atomic qubits.

\section{\label{sec:Experimental Setup}Experimental Setup}

We start the experiment by collecting $4\times 10^{9}$ $^{87}\text{Rb}$ atoms from a Zeeman slower into a magneto-optical trap (MOT). After optical pumping to the $\ket{F'=2,m_{F}=+2}$ state,  $2\times 10^{9}$ atoms are magnetically trapped in a quadrupole field with a gradient of $90$ G/cm, at a temperature of 150 $\mu$K. We characterize the ultracold atoms by fluorescence imaging in the MOT chamber, and by absorption imaging in the cryostat. Lifetimes in the magnetic trap due to collisions with the background gas in the room temperature vacuum chamber are typically 20 seconds. 

The MOT chamber is connected with a CF16 vacuum tube to the cryostat which is centered in the magnetic conveyor belt as shown in FIG. \ref{fig:setup}. The vacuum environment extends first horizontally (33.0 cm), then vertically (22.5 cm) up to the entrance of the bottom plate of a dry, UHV compatible dilution refrigerator (DR). Inside the cryostat, the atoms move further 17.5 cm vertically and pass through the 50 K shield, 3 K shield and 800 mK shield (still shield). After the still shield the atoms approach the mixing chamber plate. 

Along the vertical section, an all-metal gate valve separates the MOT and cryostat vacuum environments, to be able to detach the DR without breaking the room temperature vacuum. Before the gate valve, a 5 L ion-pump is attached to maintain the UHV inside the transport tube. The cryostat has a base temperature of 14 mK when the optical access is shielded. However when the four CF40 windows are installed the temperature raises to 70 mK, due to the additional optical heat load. Inside the cryostat the atoms are held in a quadrupole magnetic trap in proximity to the mK stage, which will later host the superconducting quantum circuit. The shield plates are equipped with mechanical shutters controlled externally in order to open and close a 12.5 mm hole for the passage of the atomic cloud. The shields are made of copper and aluminum, exhibiting very low resistivity at cryogenic temperatures, therefore the magnetic transport induces long lasting eddy currents. These eddy currents are reduced by slowing down the transport of the atoms  (see Section \ref{sec:Methods}) and by radially segmenting the plates, starting from the center hole (8 segments each plate) following the optimization using FEM simulation.

\begin{figure}[t!]
  \includegraphics[scale=0.55]{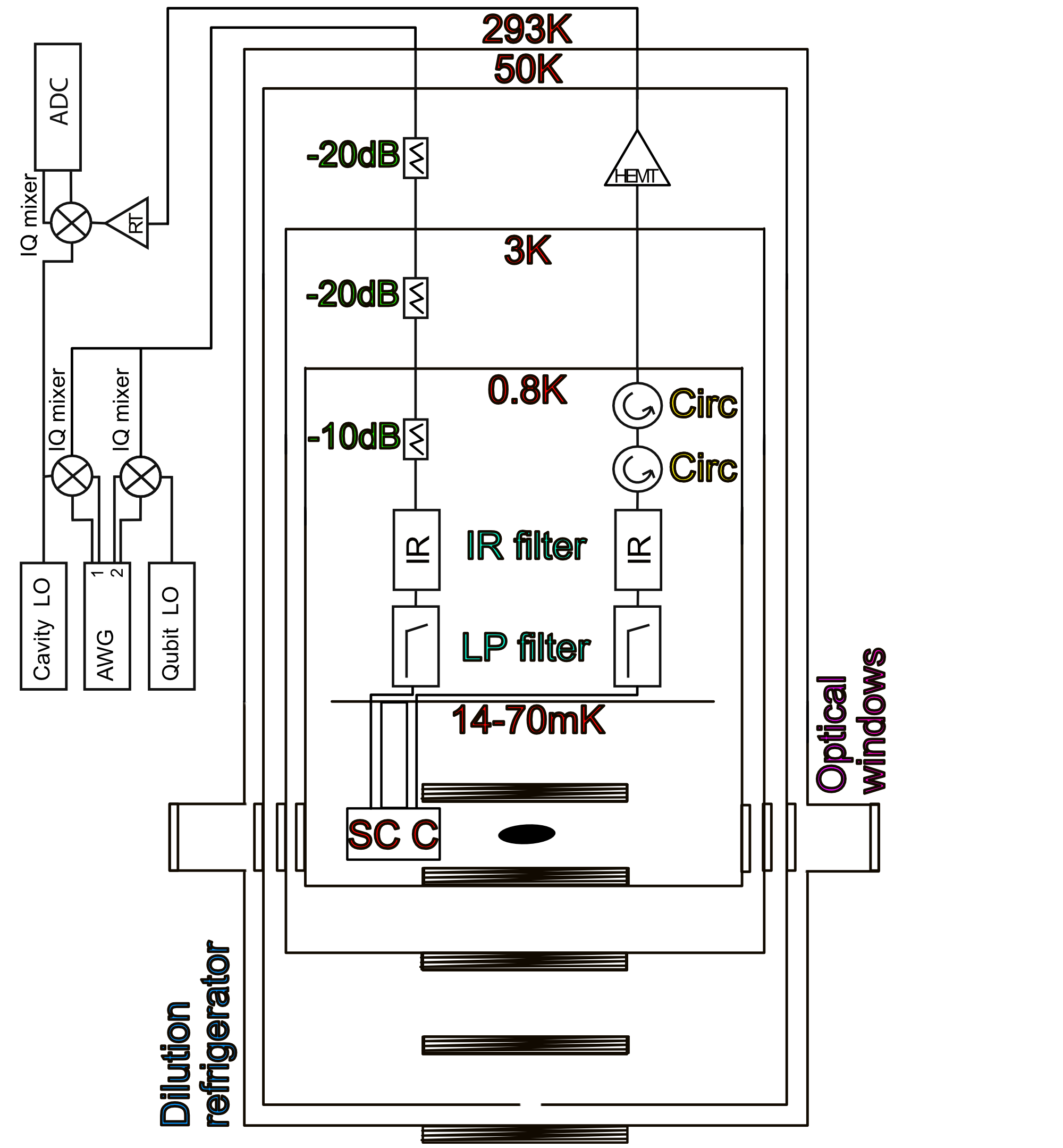}
  \caption{Schematic of the room temperature and cold microwave electronics for probing superconducting circuits. Attenuators, low pass filters (LP filters), custom made infrared filters (IR filters), circulators (Circ), HEMT and low noise room temperature (RT) amplifiers have been installed. Superconducting quantum circuits (SC C) can be probed and measured with arbitrary waveform generator (AWG), local oscillators (LO) and analog to digital converter (ADC).}
  \label{fig:setupfridge}
\end{figure}

The magnetic conveyor belt \cite{Minniberger2014} consists of 20 coils as shown in FIG.  \ref{fig:setup}. The first pair of coils coincides with the MOT coils, and the horizontal transport is achieved with four additional pairs of coils in anti-helmholtz configuration. At the 90$^\circ$ direction change, the vertical transport begins with 6 equally spaced coils that reach to the cryostat chamber. The vertical coils, casted in resin to eliminate eddy currents, have 40 windings each and are separated by 5 cm. Inside the cryostat, 4 superconducting coils, with 800 windings each, are anchored to the 3 K and still plates and wired to external current sources. The wiring consists of different sections. First the NbTi coil wire is soldered to a YBCO strip cladded in two copper layers and anchored to the shields. At the 3 K stage the YBCO strip is soldered to pre-installed HTC superconducting lines reaching up to the external room temperature feedthrough. We are able to drive 10 A on each line, without compromising the DR functionality. The room temperature coils are powered by three car batteries in series, with current regulated by a parallel MOSFET based programmable PID, connected to unipolar and bipolar switches for each coil \cite{chinchean}. The superconducting coils are driven by four dedicated bipolar current sources controlled by analog signals. 

In the cryostat stainless steel microwave lines have been installed. They are thermally anchored at each shield and they include attenuators, low pass filters, circulators, HEMT amplifier, sketched in FIG.  \ref{fig:setupfridge}. The current setup allows to probe 3D microwave cavities and SC qubits. We are currently using a rectangular cavity, in resoncance to the the atomic transition. The cavity is excited at its  $TE_{201}$ mode, allowing to have atoms and SC qubit at the magnetic and electric field antinodes respectively. A transversal through hole, 5 mm diameter, have been drilled in the cavity allowing the passage of the atoms, guided by superconducting electrodes. We have measured internal quality factors up to $Q_i=5\times 10^5$, and whenever the superconducting state is harmed by external fields, we can use laser light to reset its original status. 

\section{\label{sec:Methods}Methods}

For the magnetic transport of cold atoms, two methods are typically applied. First, a pair of moving anti-Helmholtz coils can be used to transport the atoms with the shifting quadrupole field \cite{Lewandowski2003,Nakagawa2005}. For this case the transport into a cryostat was realized in \cite{PhysRevLett.98.260407}. Second, a series of stationary, overlapping anti-Helmholtz coils can generate a moving quadrupole field by an appropriate current modulation \cite{PhysRevA.63.031401,Minniberger2014}. The latter method was preferred, as there is no moving object involved, which will be an important condition upon entry of the atomic cloud into the cryostat.

The current pulses of each coil are calculated by fixing the trap geometry, which must allow a symmetry in two directions, whereas the remaining direction defines the transport axis. To simulate the current profiles for the horizontal transport, where $x$ denotes the transport direction, we need to assume the following conditions \cite{Haslinger2011}
\begin{align}
 B(x_0) & =  0 \;\;\;\text{(at trapping position $x_0$)} \\
 \frac{\partial B_z}{\partial z} (x_0) & =  120 \text{ G/cm} \\
 A &= \frac{\partial B_y/\partial y}{\partial B_x/\partial x}(x_0)  = constant,
\end{align}
where $A$ is the aspect ratio of the atomic cloud. Contributions to the magnetic field from three pairs of coils are necessary at each point in space to meet these three conditions. For the vertical transport, four coils should be used instead, except for the start and end of the transport, fulfilling the conditions
\begin{align}
  B(z_0) & = 0 \text{\;\;\;\;(at trapping position $z_0$)} \\
 \frac{\partial B_z}{\partial z(z_0)} & = 120 \text{ G/cm} \\
 \frac{\partial^2 B_z}{\partial z^2_{z<z'}} & = 0 \text{\;\;\;\;(linear trapping gradient)} \\
 \sum_{i=1}^4 {I_i} & =0,
\end{align}
where the third condition assures the trap has a linear gradient over a wide region defined by $z'$ and the fourth requires the sum of the currents to be equal to zero, $I_i$ denoting the current in each coil.
Once the current profiles at each point of the transport have been generated, they need to be mapped into a time dependent function $I_i(x,y,z)\rightarrow I_i(t)$. While performing the mapping, care must be taken to minimize jerk, in order to prevent heating of the atomic cloud. Moreover, when the cloud experiences changes in magnetic field gradient along the path due to a mismatch in the simulated field and actual field, the cloud will heat up as well.

\begin{figure}[t!]
  \includegraphics[scale=0.51]{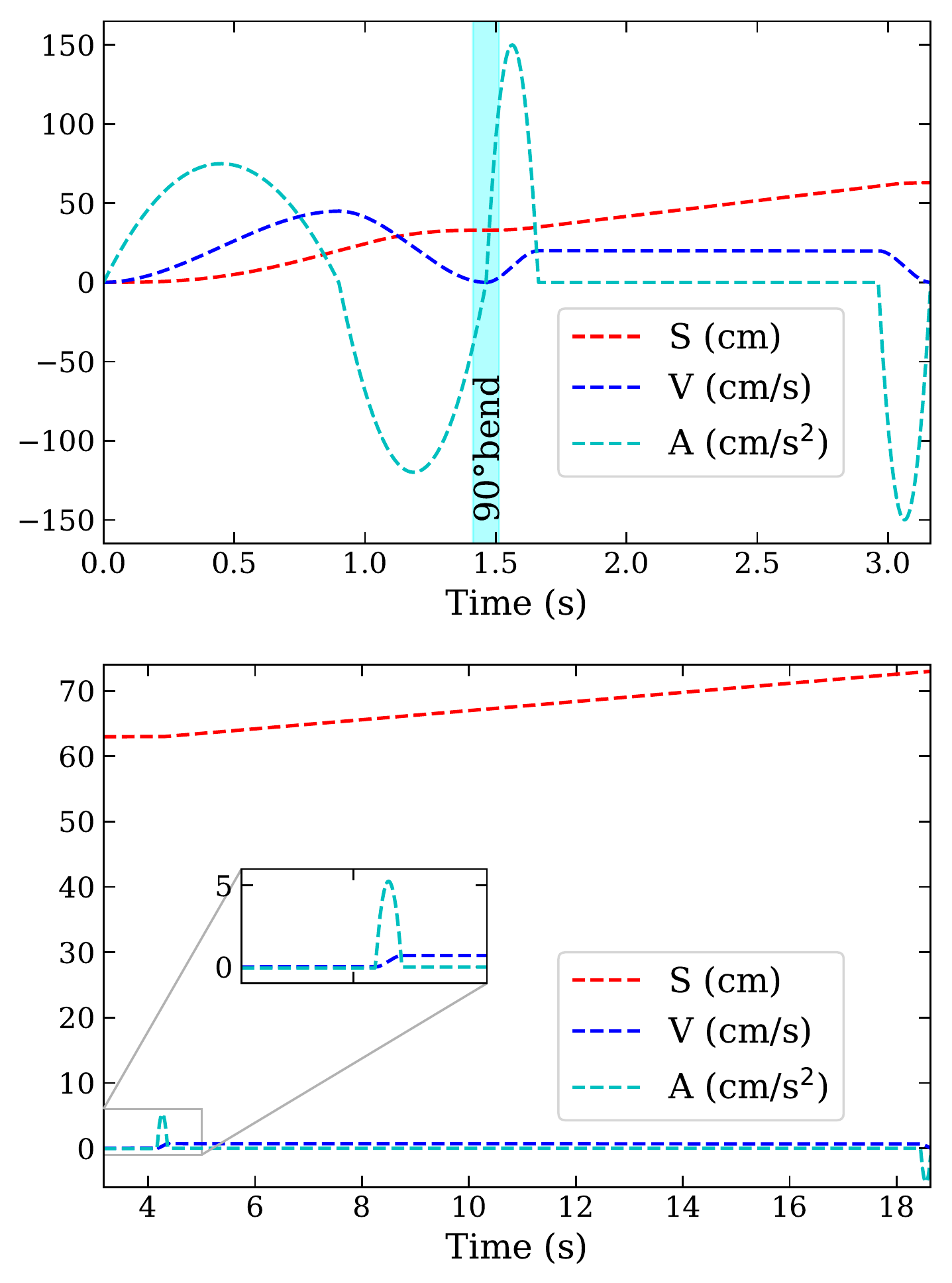}
  \caption{Acceleration, velocity and position profiles of the magnetic transport. First two sections are in the upper plot, last section is plotted below with a different timescale.}
  \label{fig:velocityprofiles}
\end{figure}

We need to define the equations of motion for space, velocity and acceleration, with the necessary boundary conditions. The whole transport is ideally divided in three distinct sections. Each of them has initial and final velocity, as well as acceleration set to zero. The first horizontal section ends at 33 cm from the MOT chamber and the optimized velocity in this segment is 50 cm/s, see FIG. \ref{fig:velocityprofiles}. After a 90$^\circ$ turn, the vertical path begins and is divided in two sections. The first vertical section extends to a total of 63 cm, of which the last 7.5 cm are inside the cryostat at 50 K temperature. The optimized maximum speed is 20 cm/s. After the first vertical section, the residual eddy currents in the segmented cryogenic shields greatly affect the transport efficiency at high velocities. This has been solved by sufficiently slowing the motion of the atoms for the remaining vertical path, benefiting from their long lifetime at cold environmental temperature (see Section \ref{sec:Results}). Therefore, in the last section which is 10 cm long (making the total vertical length to 73 cm), the atoms travel at a speed of 0.7 cm/s and have a negligible acceleration compared to the previous sections, as can be seen in the inset of the lower part of FIG. \ref{fig:velocityprofiles}. This brings the total transport duration to approximately 18 seconds. These motion profiles have been obtained through  optimization of the transport efficiency, see Section \ref{sec:Results}.
\begin{figure}[t!]
  \includegraphics[scale=0.39]{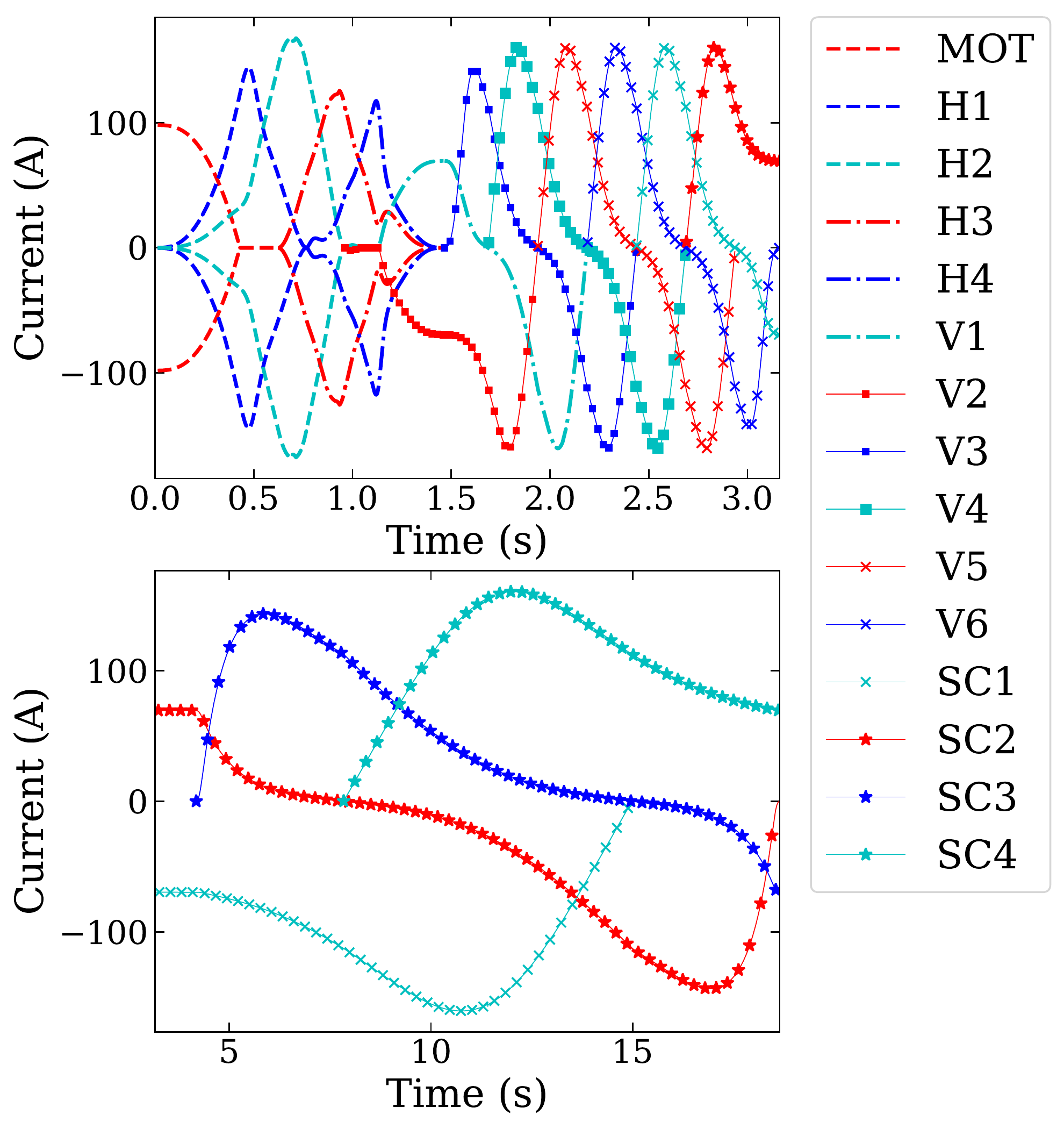}
  \caption{Actual current profiles as function of time. Last of the three transport sections is plotted separately with a different timescale.}
  \label{fig:pulseprofiles}
\end{figure}
In FIG. \ref{fig:pulseprofiles} the current pulses for each coil are plotted in time. For clarity, as in FIG. \ref{fig:velocityprofiles}, the last section has been separated due to the long duration.
\\
After the atoms have been transported in the cryogenic environment we can measure the lifetime of the atomic cloud. 

\section{\label{sec:Results}Results}

Initially, we choose a  set of equations of motion to start the optimization procedure of the transport. The cloud accelerates out of the MOT chamber into the transport tube at a distance of 20 cm. Afterwards, it decelerates for the next 12 cm of the horizontal transport.
Upon start of the transport, the magnetic trap current will be ramped down in 400 ms, while increasing the currents in the transport coils. The trap minimum will begin to move in the transport direction.
We define the efficiency parameter of the transport as $\epsilon^2=N_{t_{cut}}/ N_t$, where $N_t$ is
the atom number in the initial magnetic trap, and $N_{t_{cut}}$ is the number of atoms after back and forth transport
to an arbitrary position at the time $t_{cut}$. The returning cloud from the transport is characterized in the MOT chamber with fluorescence imaging. The efficiency is defined as the fraction of cloud remaining after the forward transport, and hence it is squared for a two-way transport.

\begin{figure}[t!]
  \includegraphics[scale=0.35]{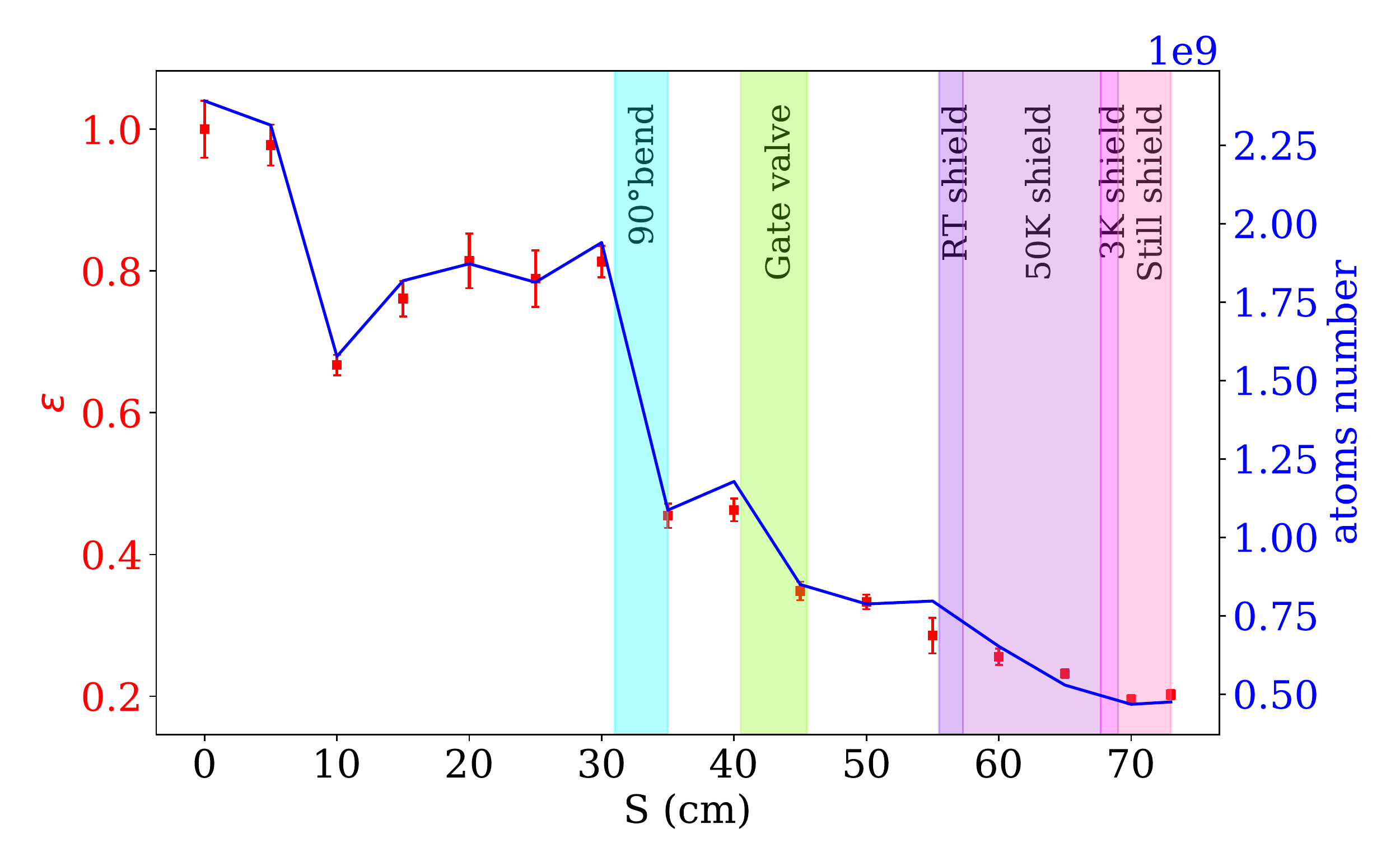}
  \caption{Efficiency $\epsilon$ of the transport on the left axis and transported atomic population on the right axis.}
  \label{fig:transportefficiency}
\end{figure}

In FIG. \ref{fig:transportefficiency} we plot the $\epsilon$ as function of transport distance corresponding to certain $t_{cut}$ values. We can observe three unusual dips in the efficiency, around 10 cm, 30 cm and 40 cm. The explanation is the following: at 10 cm we have the maximum acceleration value of the first section and by reverting the motion, for the measurement in the MOT chamber, we expect a major loss. At 32 cm we are stopping the atoms before starting the vertical transport. At this location the measured trap lifetime is 2 s and therefore the cloud experiences increased background loss. Finally, at 40 cm we have a gate valve with a residual magnetization, and although it is degaussed periodically, we still have losses there due to the distortion of the trapping magnetic field.
The final atom number at the millikelvin stage is $5\times10^{8}$. 

\begin{figure}[htbp!]
    \centering
    \includegraphics[scale=0.35]{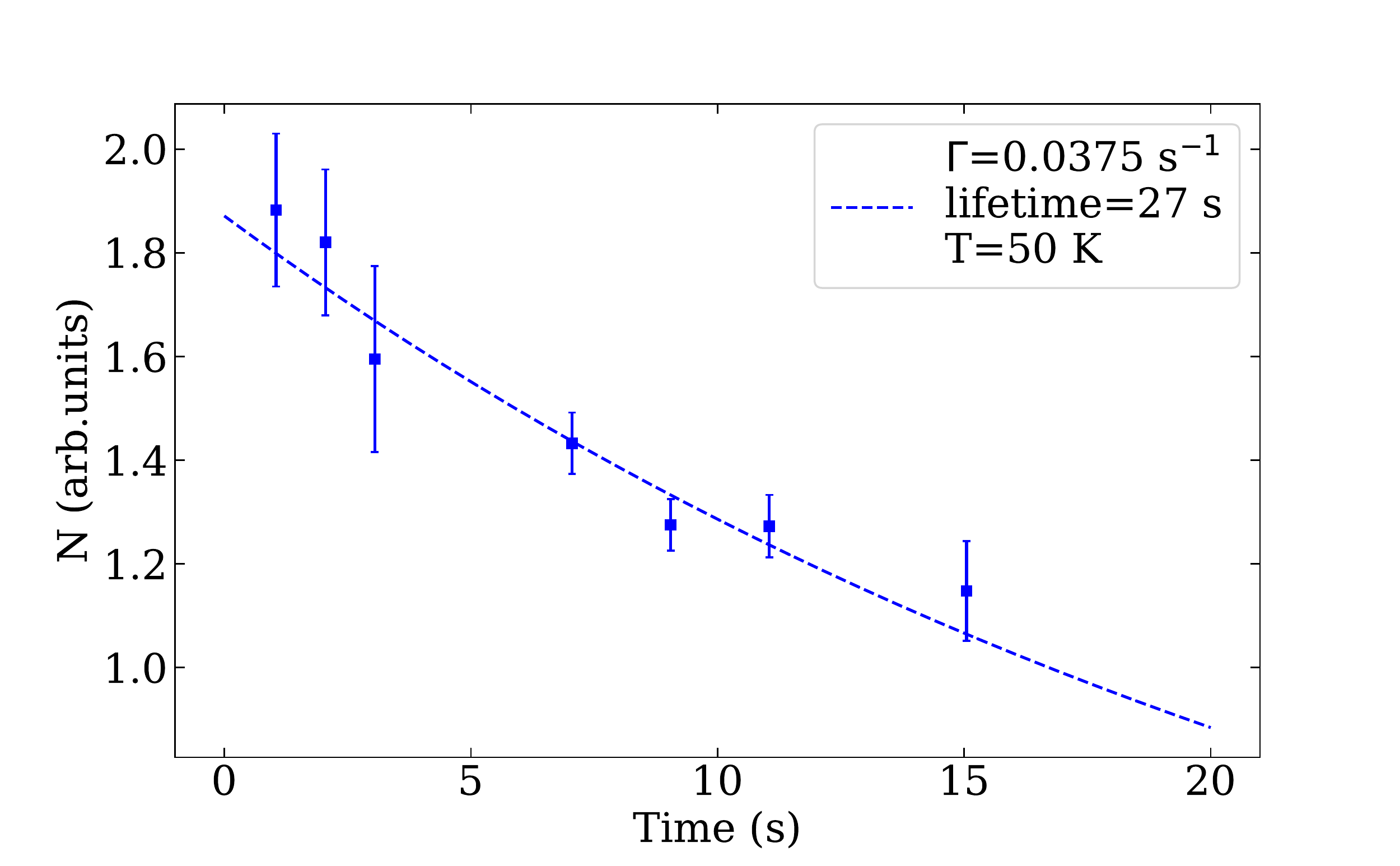}
    \includegraphics[scale=0.35]{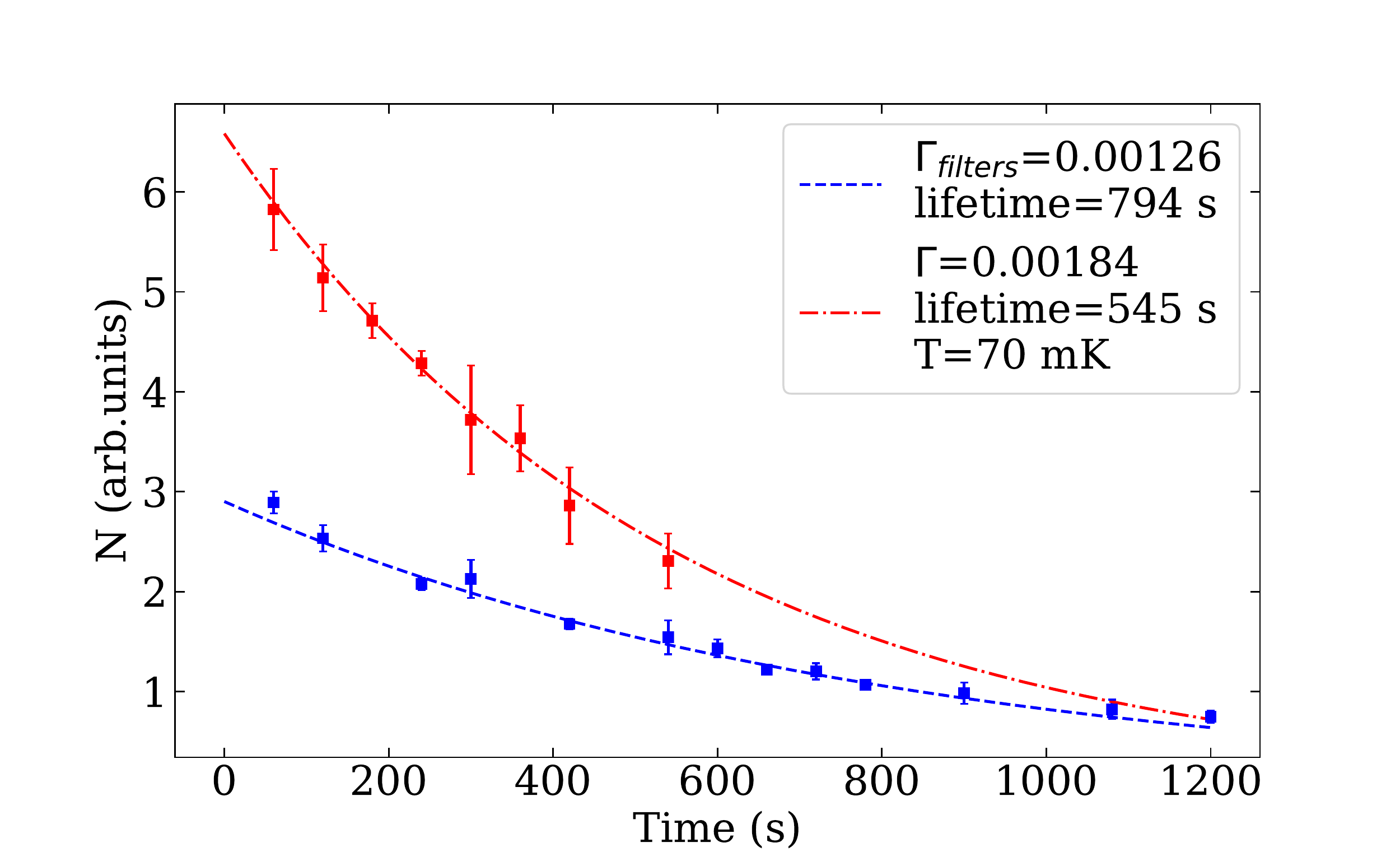}\
    \caption{$\Gamma$ decay measurements of the trap. Number of atoms in trap N, on the y-axis, is in arbitrary unit. In the upper plot, the atoms are held at the 50 K stage. In the lower one, atoms are held at 70 mK. The lifetime has been acquired with the superconducting coils current unfiltered and low pass filtered. There is a noticeable improvement when filtered.}\label{fig:lifetime}
\end{figure}

For further characterization of the cloud inside the still shield, we installed a laser beam with 260 $\mu$W power on one side of the cryostat and a CCD camera exposing 450 $\mu$s on the opposite side to perform absorption imaging. By evaluating the atom number for increasing holding times in the trap, one can estimate the lifetime from the inverse of the exponential decay $\Gamma$ in FIG. \ref{fig:lifetime}. The measurements have been acquired at the 70 mK stage through the viewports and at the 50 K stage by reverting the motion of the atoms back into the MOT chamber, since there are no windows at this stage. The lifetime at the 50 K stage is 27 s, limited by the background pressure. At the 70 mK stage the lifetime increased to 13 minutes.

We note that the lifetime was enhanced from below 10 minutes to 13 minutes by applying a low-pass filter in the drive of the coils  (R=1 $\Omega$, C=30 $\mu$F, $f_{cut}$=5 kHz) to prevent electrical noise. In addition, we improved the vacuum pressure between the room temperature shield and the 50 K shield by installing non evaporable getters and baking the outer shield of the cryostat prior to cooldown.  We  expect a negligible desorption from the millikelvin surfaces, which includes the 70 mK plate and the still shield. Furthermore, with closed shutters and viewports light-induced atom desorption is not present, as suggested as a  limiting factor in \cite{PhysRevA.51.1403}. 

From the decay $\Gamma$ of the atomic cloud, the estimated pressure inside the still shield is $P=2.55\times 10^{-13}$ mbar. We report the model of the elastic collisions in the  appendix.

\section{\label{sec:Conclusions}Conclusions}

In conclusion, we demonstrated an experimental platform for hybrid quantum systems, which is capable of combining ultracold atom and superconducting circuit physics in a single setup. We showed that we can routinely transfer clouds of $5\times10^8$ atoms close to the mK stage of a dilution refrigerator, at a base temperature of 70 mK. Lifetime measurements of the atomic cloud inside the cryogenic environment showed values of 13 minutes, which is a record in atomic physics experiments. This  long lifetime gives us access to employ the ultracold atoms as an extremely sensitive probe for detecting static and fluctuating electric and magnetic fields. In future, we intend to use this setup to study interfaces between ultracold atoms and superconducting circuits. Different schemes for the merging and coupling of both systems are in preparation. 

\subsection*{\label{sec:AppendixA}Appendix: Scattering theory of background gas collisions}

The elastic collisions between the alkali atoms and the surrounding gases make the atoms escape the trapping potential, defining the lifetime. The loss coefficient can be generally expressed as \cite{PhysRevA.60.R29,PhysRevA.84.022708,Eckel_2018}
\begin{equation}
    \Gamma=\sum_{i}n_{i}\langle\sigma v\rangle_{X,i},
	\label{lossrate}
\end{equation}
where $n_i$ is the density of the $i_{th}$ background species, $X$ is the trapped species and the velocity-averaged loss cross section $\langle\sigma v\rangle$ is a function of the trap depth $U_{trap}$. If we consider the elastic scattering of an alkali atom and a scattering particle, the change in kinetic energy of the atom is defined by
\begin{equation}
    \Delta E\simeq\frac{\mu^2}{M_a}\lvert\vec{v}_r\rvert^2(1-\cos{\theta}),
	\label{kineticenergy}
\end{equation}
where $\mu$ is the reduced mass of the system, $M_a$ is the mass of the atom alone, $\theta$ the collision angle, $\vec{v}_r=\vec{v}_a-\vec{v}_b$ is the initial relative velocity between the particles, with $b$ being the background species. $\Delta E$ exceeds the trap depth for a minimum angle $\theta$
\begin{equation}
    \theta_{min}=\arccos(1-\frac{M_a U_{trap}}{\mu^2 \lvert\vec{v}_r\rvert^2}).
	\label{thetamin}
\end{equation}
The differential scattering cross-section is defined as $d\sigma/d\Omega$, and it is equivalent to the quantum-mechanical scattering amplitude $\lvert f(k,\theta)\rvert^2$, by using the continuity equation for the wavefunction and the probability current density.  For an incident scattering particle with wave number $k$, the cross section for loss-inducing collision from a trap of depth $U_{trap}$ is 
\begin{equation}
    \sigma_{loss}=\int_{\theta_{min}}^\pi 2\pi \sin{\theta}\lvert f(k,\theta)\rvert^2 d\theta.
	\label{sigmaloss}
\end{equation}
By solving for the velocity average over the Maxwell-Boltzman distribution one can obtain \cite{PhysRevA.84.022708,VanDongen_2014}
\begin{equation}
    \langle\sigma v\rangle=\left(\frac{M_b}{2\pi k_b T}\right)^{3/2}\int_{0}^\infty 4\pi \sigma_{loss}(k)v_{b}^3 e^{-M_b v_{b}^2/2k_b T}dv_b,
	\label{scattcross}
\end{equation}
where it is assumed that the atom is steady, $\vec{v}_r\simeq\vec{v}_b$, and $k=\mu v_b/\hbar$ and $\sigma_{loss}$ contains the spherical harmonics.\\

It can be seen in \cite{PhysRevA.85.033420,PhysRevA.38.1599} that we can further simplify the problem obtaining 
\begin{equation}
    \gamma_{i} \approx 6.8\frac{P_i}{(k_B T)^{2/3}}\left(\frac{C_i}{M_b}\right)^{1/3}(U_{trap}M_a)^{-1/6},
	\label{gammai}
\end{equation}
where $\Gamma=\sum_{i}\gamma_{i}$, $C_i$ the species dependent Van der Waals coefficients estimated with the Slater-Kirkwood formula, which for Helium, it is equal to 35 in atomic units \cite{PhysRevA.85.033420,MARGENAU196915,MILLER19781}. $P_i$ represents the partial pressure, $T$ the background gas temperature and mass $M_b$ and $M_a$ the rubidium mass. We must note that the trap depth $U_{trap}$ varies with the laser and magnetic trap parameters and  can  be  challenging  to  quantify, usually varying between 0.5 and 2 K \cite{PhysRevA.84.022708,PhysRevA.56.4055}, but with a weak dependence in $\gamma_{i}$. At the considered temperatures, only Helium atoms contribute to the background gas vapour pressure.  Following the above considerations, one can estimate the parameter $\Gamma/P=4.93\times 10^9$ mbar$^{-1}$s$^{-1}$ for collisions between ground-state Rb atoms and Helium-4 at a background temperature of $T=0.07$ K and $U_{trap}\approx 1$ K.
Using Eq. \ref{gammai} and substituting the exponential decay $\Gamma$ of the cloud in the magnetostatic trap in the cryostat from FIG. \ref{fig:lifetime}, we obtain a pressure of $P=2.55\times 10^{-13}$ mbar, below the one reported at 3.6 K temperature in \cite{PhysRevA.51.1403}, which is with nearly 10 minutes lifetime.

%\bibliography{report} % bibliography data in report.bib

\bibliographystyle{apsrev4-2} 
\end{document}